\documentclass[preprint2]{aastex}

\usepackage{url}

\shorttitle{Fundamental-Mode Oscillations of Two Coronal Loops}
\shortauthors{Jain et al.}

\begin{document}
\title{Fundamental-Mode Oscillations of Two Coronal Loops within a Solar Magnetic Arcade}

\author{Rekha Jain, Ram A. Maurya}
\affil{School of Mathematics and Statistics, University of Sheffield,
    S3 7RH, UK}
\email{R.Jain@sheffield.ac.uk}

\author{Bradley W. Hindman}
\affil{JILA, University of Colorado, Boulder, CO~80309-0440, USA}

\begin{abstract}

We analyse intensity variations, as measured by the Atmospheric Imaging Assembly (AIA) in the 171 {\AA} passband, in two coronal loops embedded within a single coronal magnetic arcade. We detect oscillations in the fundamental mode with periods of roughly 2 minutes and decay times of 5 minutes. The oscillations were initiated by interaction of the arcade with a large wavefront issuing from a flare site. Further, the power spectra of the oscillations evince signatures consistent with oblique propagation to the field lines and for the existence of a 2-D waveguide instead of a 1-D one.

\end{abstract}

\section{Introduction and Background}

Bright coronal loops are often observed to sway back and forth (e.g, Aschwanden et al. 1999). Usually, the oscillations begin suddenly in response to a nearby flare or filament eruption (Wills-Davey \& Thompson 1999). After initiation the amplitude of the motion attenuates, disappearing below the noise threshold after a few wave periods (Nakariakov et al. 1999; White \& Verwichte 2012; Nistico et al. 2013). These oscillations have been studied intensely because of the diagnostic potential they offer for measuring the magnetic properties of the loops upon which they reside (Roberts 2000; Andries et al. 2009). However, before such diagnostics can be fully exploited, a detailed understanding of the nature of the waves and the waveguide is required (Jain \& Hindman 2011). For instance, all of the seismic inferences that have made to date about loop properties have implicitly assumed that the observed oscillations are standing waves. Unfortunately, verifying that the observed oscillations are not the response to a transient travelling wave is difficult and only a few studies have successfully done so (Aschwanden \& Schrijver 2011; Verwichte et al. 2004). Further, recent work by Hindman \& Jain (2014) has even called into question the 1-D nature of the loop waveguide that has been the theoretical paradigm. They suggest that the waveguide may be comprised of the entire magnetic arcade in which the visible loop is embedded.

The exact mechanism by which a flare induces oscillations in nearby loops is not fully understood at present; however, Hudson \& Warmuth (2004) have suggested that the magnetic blast wave that is expelled from the flare site may be important. Observations connecting loop oscillations with such expanding disturbances have been reported in the past (e.g., Wang et al. 2012). The fact that some loop oscillations appear to grow in amplitude initially before reaching maximum strength before a possible decay, indicates that the excitation event may be temporally extended and that not all loop oscillations are excited in the same manner. Furthermore, the polarisation could also depend on the loop oscillation plane (see Wang and Solanki, 2004). The swaying motions transverse to the loop plane are referred to as {\it horizontal oscillations} whereas the motions polarised in the loop plane can either conserve the loop shape ({\it vertical oscillations}) or length ({\it distortion oscillations}).

Here we report on observations of two distinct coronal loops that were both part of the same magnetic arcade. Both loops began oscillating after the passage of an expanding wavefront that spread from the site of a nearby flare. In section 2, we describe the data employed, depict the geometry of the loops, and discuss the timing of the wavefront and the commencement of the oscillations. In section 3, we present a time-series analysis of the loop oscillations, power spectra and cross-correlations. In so doing, we provide clear evidence that the oscillations were standing waves in the fundamental mode. Finally, in section 4 we discuss the implications and importance of our findings.

\section{Data}

We considered EUV images of NOAA AR1283 with pixel resolution of 0.6$^{''}$ taken by the Atmospheric Imaging Assembly (AIA) on board the Solar Dynamics Observatory (SDO) in the 171 {\AA} passband (Lemen 2012) on 6 September 2011. The dataset was chosen for studying coronal-loop oscillations which were observed immediately after the passage of a wavefront launched from the flare site. A few distinct loops within a magnetic arcade were visible and were seen to oscillate.

Figure 1a displays AR1283 and the area indicated by dotted lines is featured in a zoom-in view in Figure 1b. Figure 1c presents an overlay of the same region for three different AIA passbands: 171 {\AA}, 193 {\AA}, and 221 {\AA}. The flare occurring in AR1283 was an X2.1 class flare located close to the disk centre at latitude 14$^{\circ}$ North and longitude 18$^{\circ}$ West. It initiated at 22:12 UT, peaked at 22:20 UT, and ended at about 22:30 UT. Figure 1d shows the fluxes in GOES 4 {\AA} and 8 {\AA} as well as the AIA 171 {\AA} passband. The flare was also associated with a coronal mass ejection (CME) (see for example, the SOHO/LASCO CME catalogue). The metric type burst II was not obvious in one band for this event but was reported to be present in full range (see, GBSRBS\footnote{\url{http://www.astro.umd.edu/~white/gb/Html/Events/TypeII/20110906_221400_merge.html}}). Note that EUV dimming in AIA 171 {\AA}, shown in Figure 1d is also an indication of a launched CME.

The wavefront ejected from the flare was clearly seen in the AIA 221 {\AA} passband (see Figure 2a). By tracking the position of this wavefront, we estimate a propagation speed of 870 km s$^{-1}$. Careful investigation suggests that the propagating wavefront was visible just before the maximum in the release of flare energy as measured by GOES. The oscillations were initiated at the same time that the wavefront struck the loops. Figure 2b illustrates this point, showing an AIA 171 {\AA} image with the location of the wavefront indicated for two different times. The footpoints of the loops (located near the left end of the wavefront arcs) are impacted around the time 22:18 UT, which, as we will see later, is the time at which the loops began to oscillate.

From the AIA imagery it appears that the oscillating loops were part of a low-lying magnetic sheet which was arched over by a higher set of upper loops with a slightly different orientation. When the wavefront hit the entire structure, the upper loops lifted and as they did so, their rightmost portions disappeared and may have detached. This is most easily seen by examining the supplemental online movie\footnote{\url{http://www.rekha-jain.staff.shef.ac.uk/Suppl_mat/overlying_loops.mp4}}. Since the passage of wavefront, the readjustment of the geometry of the upper loops, and the start of oscillations on the lower loops all occurred simultaneously, it is impossible to know the direct cause of the oscillations. The wavefront could have shaken the oscillating loops directly or the sudden lifting of the upper field lines may have induced a pressure fluctuation that set the lower loops oscillating.  But since the lifting of the upper loops seems to have been instigated by the passage of the wavefront, the wavefront probably played at least an indirect role.

Figure 3 shows snapshots of the magnetic arcade at two different times spanning the event. The top panels, (a) and (c), show the location of two loops that we will study in detail. These loops are labelled A and B. Since only a fraction of each loop is seen by STEREO-A, we reconstructed the loop orientation by modelling the shape of the loops with an inclined semi-circle, as suggested in Aschwanden et al. (2002). With this procedure we obtained loop lengths of 158 Mm and 164 Mm. Diagrams of the loop geometry and their projections on the plane of the sky are presented in the right panels of Figure 3.

\section{Time-Series Analysis}

To study the oscillations we extracted the temporal variation of the AIA 171 \AA\ intensity along the eight slits whose positions are indicated in Figure 3b. The intensity for each slit is illustrated as a time-distance image in Figure 4. The slit numbers are marked in the top left corner. The oscillation of both identified loops are particularly evident in slit 7. The oscillations begin around 22:18 UT, which corresponds with the passage of the wavefront launched from the flare.
Figure 5a shows the intensity for a segment of slit 7 centered on the oscillating loops. By fitting a parabola locally to the intensity of each loop separately, we find the position of peak brightness as a function of time and generate a time series for the position of each loop in the slit. These time series are shown with the red and blue crosses in Figure 5b. Each time series has the appearance of a decaying sinusoid. Therefore, we fit each with a function of the form ${\cal A}\ \exp(-t/\tau)\cos⁡(2\pi t/T +\phi)$, where ${\cal A}$ is an amplitude, $\tau$ is a damping time, $T$ is a wave period, and $\phi$ is a phase. These resulting fits are indicated in Figure 5b with the dotted curves. The phase difference between the two time series is 31$^{\circ}$ and both have a period of roughly $T=2$ min and a decay time of $\tau=5$ min. We note that oddly the phase of loop A lags that of B, despite the fact that loop A is closer to the flare. 

\subsection{Power Spectra}
The power spectra for all the time series (data and fits) are shown in Figure 5c. The dataset has a Nyquist frequency of approximately 21 mHz and a frequency spacing of 1.4 mHz. The oscillations of both loops have a single dominant frequency of roughly 8 mHz and are consistent with a lifetime of 5 min (i.e., a half-width of 0.5 mHz). However, both of the spectral profiles for the unfitted data are asymmetric with enhanced power in the high-frequency wing. Such a result is expected when modelling the loop oscillations as the response of a magnetic arcade to fast MHD waves. The waveguide is 2D and the waves are standing oscillations in the direction parallel to the field lines and may be travelling disturbances perpendicular to the field down the axis of the arcade. The power peak in this 2-D arcade model is due to those waves that propagate parallel to the field lines, while the enhanced power in the high-frequency wing is the contribution from waves that propagate both parallel and transverse to the field lines. Since only one power peak is observed, we believe that the oscillations correspond to the fundamental resonance of the waveguide with the wavefunction lacking internal nodes between the fixed footpoints.

\subsection{Phase variation along loop}

We verify that the oscillation is indeed the fundamental waveguide mode by performing a careful analysis of the phase of the oscillation along the loop. For loops A and B, we generate time series for the oscillations for each slit by repeating the previous analysis. Subsequently, we cross-correlate the time series for two different slits and find the time lag that corresponds to the centroid of the correlation peak. The results are illustrated in Figure 6, which shows the cross-correlation as a function of time lag between the time series for slit 7 and the series for every other slit. The resulting time lag between the signals is uniformly small $\Delta t <$ 5 s, consistent with zero lag to within the observational errors (the temporal cadence of the interpolated time series is 12 s). Thus the entire loop oscillates in phase, indicating that the wave is a standing wave and further, the loop is oscillating in the fundamental mode.

\begin{figure*}
\begin{center}
\includegraphics[width = 0.6\textwidth,clip=,bb=25 13 431 360]{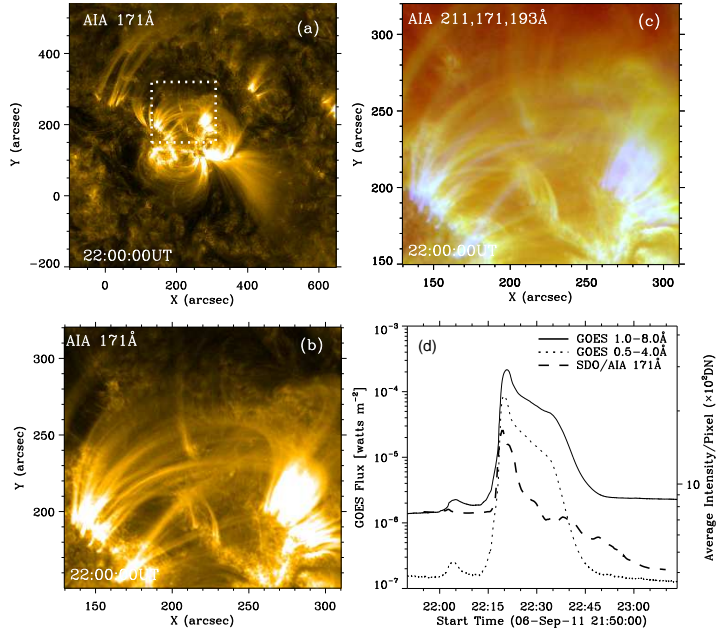}
\caption{Snapshots of NOAA AR1283. (a) The region of the flare and the magnetic arcade as seen in AIA 171 {\AA}. (b) Zoom-in view of the area shown in panel a. (c) The same region shown in an overlay of AIA 171 {\AA}, 193 {\AA}, and 211 {\AA}. (d) Energy flux due to the flare as seen in AIA 171 {\AA} and the GOES wavelengths.}
\end{center}
\end{figure*}

\begin{figure*}
\begin{center}
\includegraphics[width=0.4\textwidth,clip=,bb=8 16 247 246]{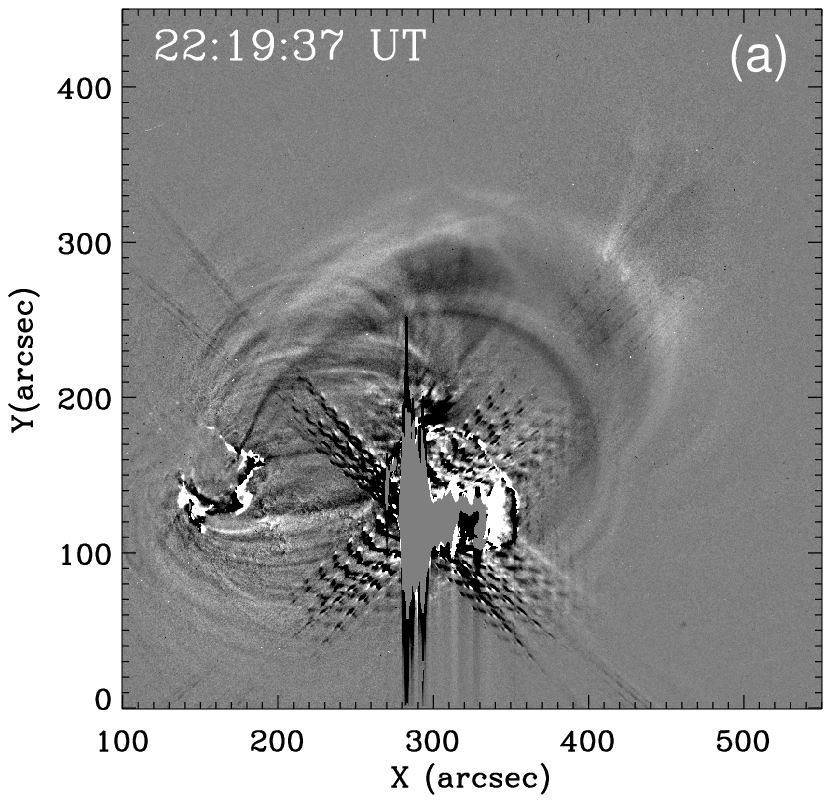}
\includegraphics[width=0.4\textwidth,clip=,bb=8 15 246 247]{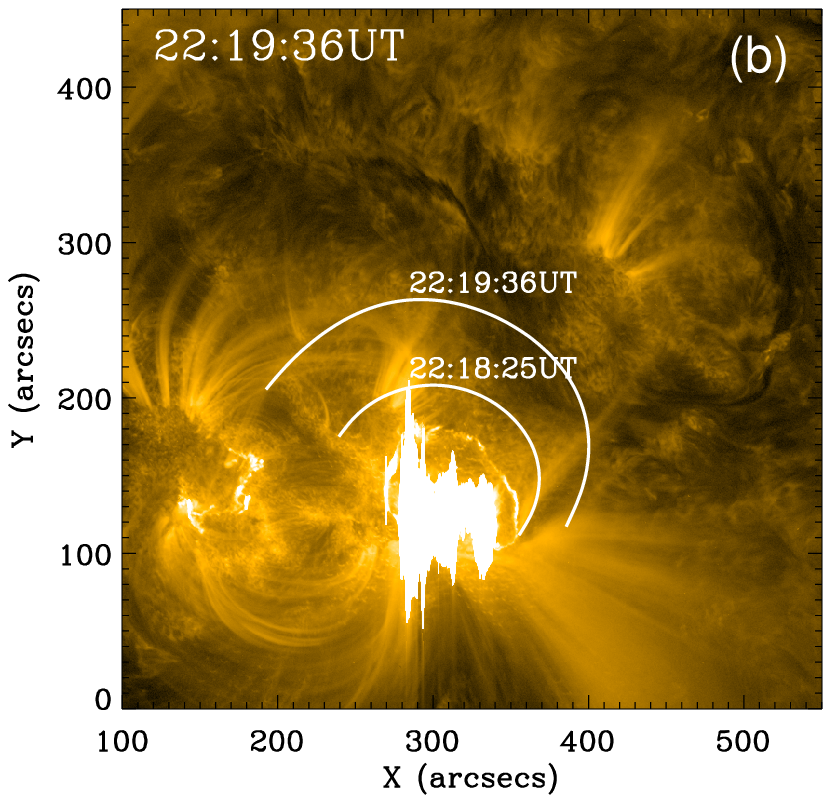}
\caption{The flare as seen in by AIA. (a) Wavefront launched from the flare as seen in 211 {\AA} passband. (b) Location of the wavefront projected on AIA 171 {\AA} image to show the interaction of the wavefront with the loop and the onset of the oscillations.}
\end{center}
\end{figure*}

\begin{figure*}
\begin{center}
\includegraphics[width=0.4\textwidth,clip=,bb=35 38 234 393]{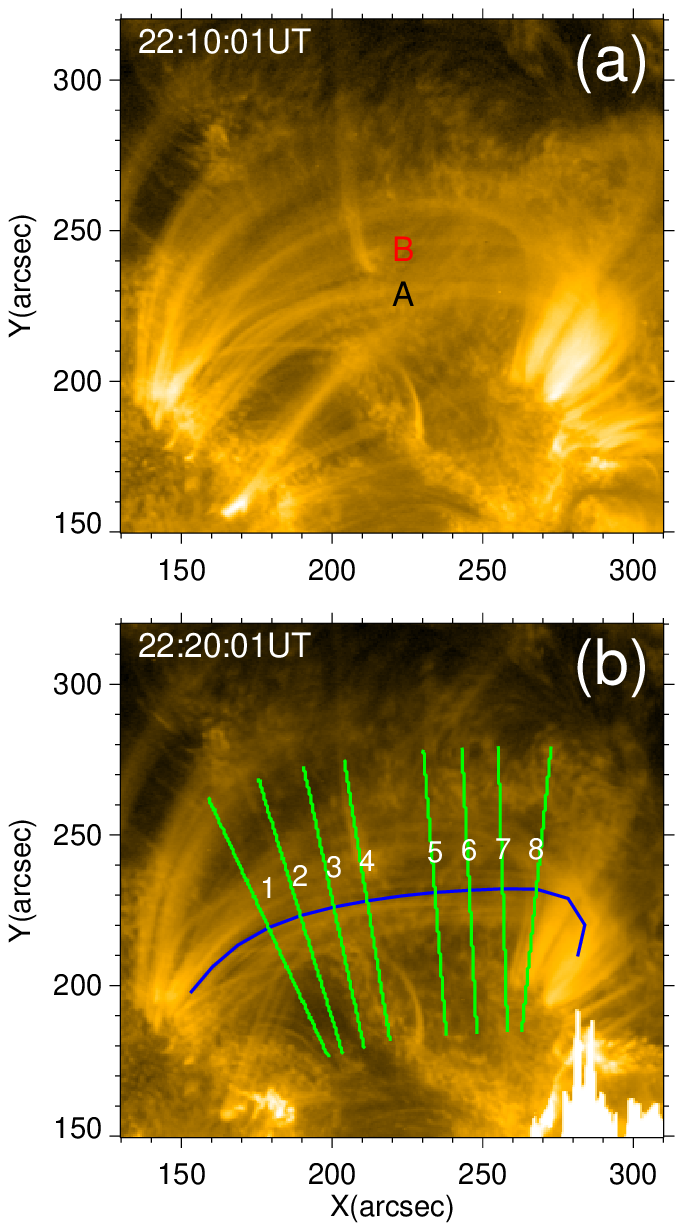}
\includegraphics[width=0.36\textwidth,clip=,bb=7 16 210 409]{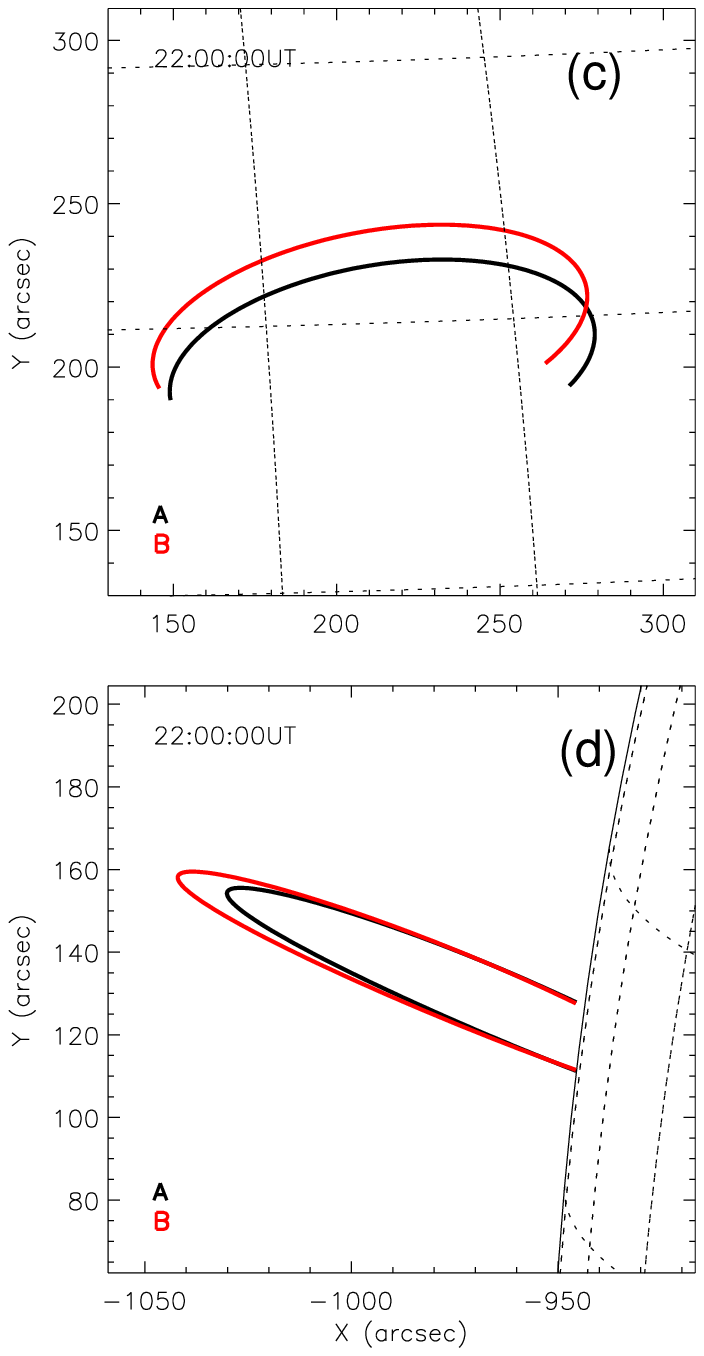}
\caption{Geometry of the two coronal loops that were seen to oscillate. (a) Snapshot of the arcade at a time before the flare. The two loops are labelled A and B. (b) The same arcade as viewed at the time of the flare. The position of loop B is indicated in blue. The location of eight slits used in the time-series analysis are shown with the green lines. (c) The reconstruction of the full loops by modelling each as an inclined semi-circle, as seen from the same vantage as the AIA instrument. (d) The projection of the loops as they would be seen at the limb.}
\end{center}
\end{figure*}

\begin{figure*}[ht]
\begin{center}
\includegraphics[width=0.6\textwidth,clip=,bb=20 0 453 410]{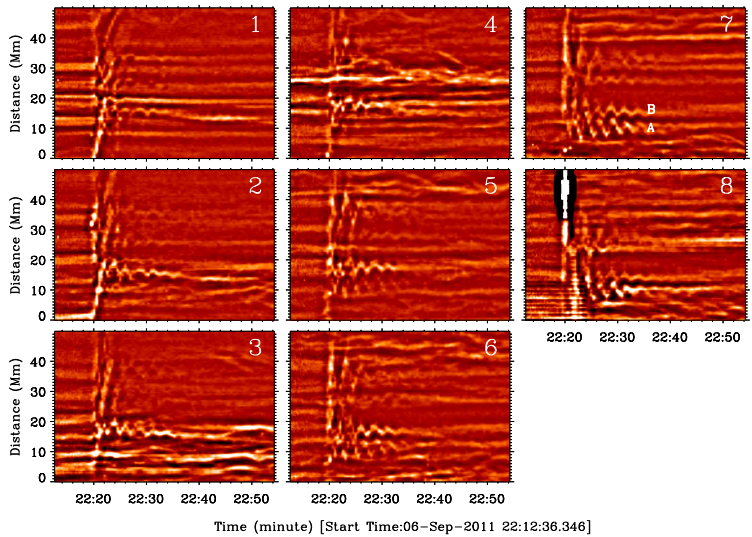} 
\caption{Smoothed time-series of the intensity in AIA 171 {\AA} for each slit shown in Figure 2c.}
\end{center}
\end{figure*}
 
\begin{figure*}
\begin{center}
\includegraphics[width=0.4\textwidth]{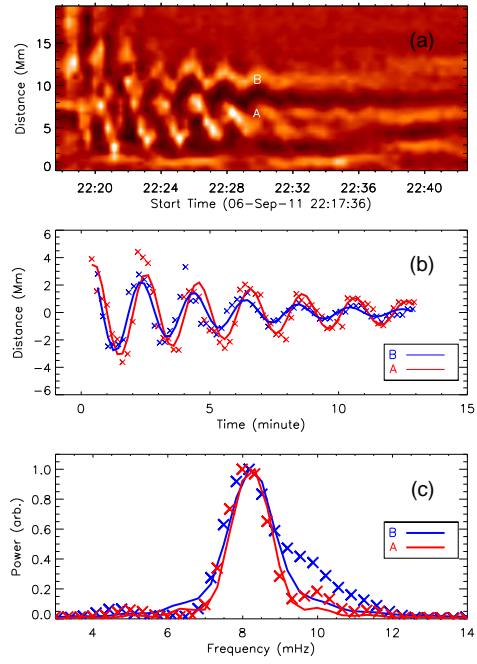} 
\caption{Time series analysis for the intensity observed in slit 7. (a)\ Time series of the intensity in AIA 171 {\AA} for loops A and B. (b)\ Symbols represent the position of maximum intensity for loops A and B. The solid curves are the fits with a damped sinusoid (see the text). (c)\ Normalised power spectra of the time-series: symbols for the data represented by symbols in panel (b) and solid lines for the fitted time series in panel (b).}
\end{center}
\end{figure*}
 
\begin{figure*}
\begin{center}
\includegraphics[width=0.52\textwidth]{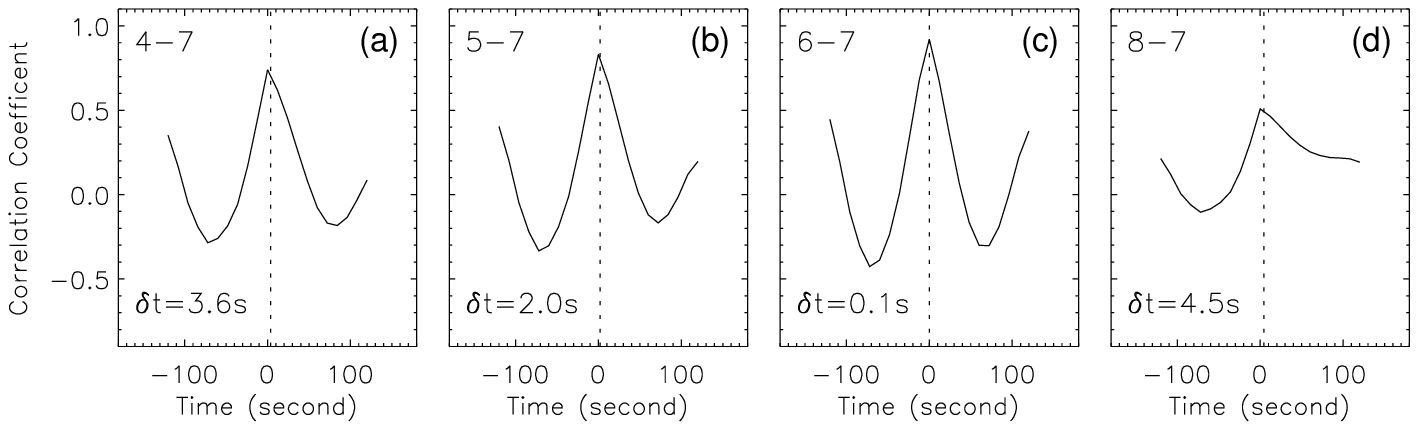}
\includegraphics[width=0.52\textwidth]{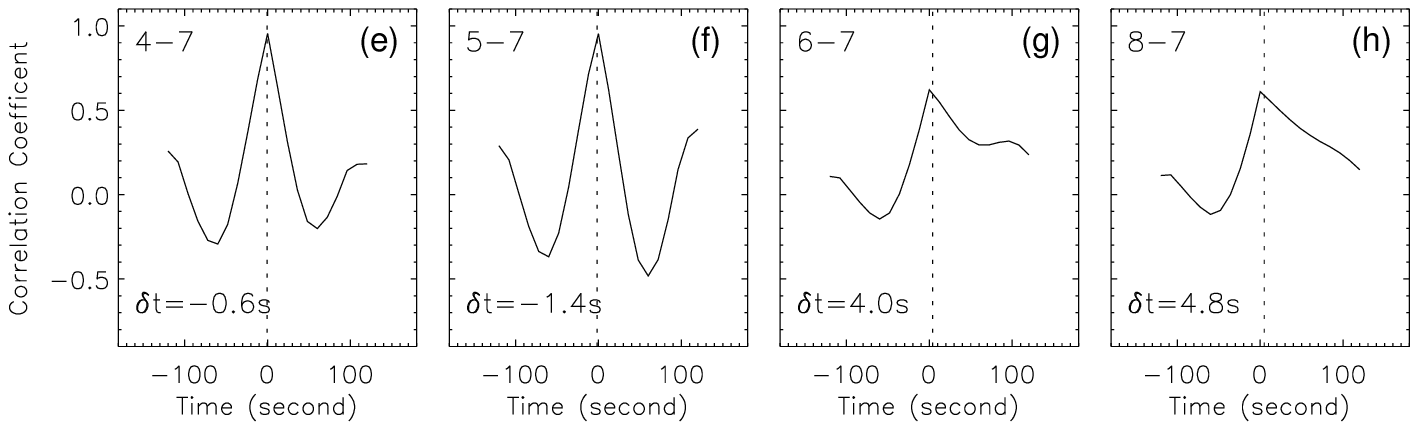}
\caption{Correlation coefficient as a function of time lag for slits on either side of slit 7 (e.g., correlation between slit 4 and slit 7 is denoted in the top right corner by 4-7). The top panels, (a)--(d), are for loop A and the bottom panels, (e)--(h), are for loop B.  All positions along a loop oscillate in phase, without a significant phase difference, indicating that the waves are the fundamental mode.}
\end{center}
\end{figure*}

\section{Discussion and Summary}
We performed a time-series analysis of the oscillations of two coronal loops that were embedded within a single magnetic arcade. Both loops had a similar length of $\sim$160 Mm, but slightly different orientations (see Figure 3). The observed oscillations were triggered by a nearby X2.1 flare and the initiation time of the oscillations is well-correlated with the passage of a wavefront launched from the flare site. Although the event was associated with a coronal mass ejection (CME), there was only a partial opening of field lines in the arcade. There was no significant restructuring of the oscillating coronal loops themselves.

The power spectra for the oscillation of both loops clearly show a single {\sl asymmetric} peak as a function of frequency with a high ratio of signal to noise of roughly 10. The power peaks at a frequency of 8 mHz corresponding to a period of 2 min. The oscillations have an amplitude that attenuates with an e-folding time of 5 min.  This time scale is well-matched by the half-width of the power peak when measured on the low-frequency side of the peak. We analyzed the time series of oscillations at eight different slit positions along the loops. Through cross-correlation of the signal from one slit with that of another, we found that the full length of the loop oscillates in phase. The single power peak and the phase measurements unambiguously demonstrate that the observed wave was a standing wave parallel to the field lines and that only the fundamental mode of the wave cavity was excited. From the frequency of the oscillations and the loop lengths, we can estimate that the mean wave speed along the loops. Using $v = 2L/T$ with loop-length, $L = $ 160 Mm and the time period, $T\approx$ 130 s, the wave speed $v$ is $\sim$ 2500 km s$^{-1}$.

Both loops reveal the unusual feature that the high-frequency wing of the spectral power profile was enhanced compared to the low-frequency wing (see Figure 5c). Such behavior was previously predicted by Hindman \& Jain (2014). They argued that some coronal-loop oscillations are the response of the entire magnetic arcade to an MHD fast wave disturbance. The arcade acts as a  2-D waveguide, for which the wavemodes are standing waves in the direction parallel to the field lines and are trapped between the footpoints. Simultaneously, these wavemodes are free to propagate perpendicular to the field lines in the direction parallel to the axis of the arcade. This model predicts that each mode of the waveguide should add its own individual asymmetric power peak to the spectrum. The core of the power peak is formed by those waves that propagate almost parallel to the field lines and the high-frequency enhancement of the power wing is caused by the existence of waves that propagate obliquely and bounce back and forth between the footpoints as the wave travels down the arcade. While our observed spectrum is well-matched by such a model, our evidence is incomplete. Clear proof would have been a specific phase relation between different loops within the arcade. This phase relation is not reproduced and in fact an examination of Figure 5 reveals that the loop farthest from the flare precedes the closer loop in phase. However, these phase arguments were predicated on the assumption that wave excitation occurs suddenly at a single location within the waveguide; thus, the phase difference between loops depends only on the travel time between loops. The situation observed here is far more complex. The wave source is not necessarily sudden and it certainly isn't spatially compact. We suggest two possible excitation mechanisms: direct shaking of the oscillating loops by the wavefront that expands from the flare site and the sudden reduction in pressure caused by the rise of the upper field lines.

While the wavefront is a spatially narrow feature, it still should cause temporally and spatially distributed driving. The wavefront takes a finite time to cross an individual loop, roughly 100 s given the speed of the wavefront, the orientation of the loops, and the distance between footpoints (roughly 95 Mm). Further, the wavefront crosses the entire arcade potentially exciting waves all along its length. In this case, the absolute phase of the oscillations are not an indication of a single excitation time, but instead they depend on the complicated details of the wavefront's passage.

Another possibility is that the upward lifting of the overlying field lines caused a pressure fluctuation that induced waves on the lower loops. In this case, the driving would be fairly uniform along the length of each loop; thus, preferentially exciting the fundamental mode of oscillation as observed. Further, since the upper loops cross over the lower loops and they appear to have detached from their right footpoints, the lower loops most distant from the flare may have been excited first as the disturbance travelled from the right footpoints to the left along the upper loops. While conjecture, this might be the simplest explanation for the puzzling fact that the most distant loop was observed to have the most advanced phase. Considering the separation between the two loops (roughly 3.5 Mm) and the measured time lag of 10 s between the two loops (31 degree phase difference with a period of 2 min), the wavespeed in the upper loops would need to be on the order of 350 km s$^{-1}$.

We have studied oscillations of two distinct coronal loops that were part of the same magnetic arcade. These loops started to oscillate after the passage of an expanding wavefront that spread from the site of a nearby flare. A time-series and phase analysis of the loop oscillations suggest that the oscillatory motions are fundamental modes of oscillations with 2-minute periods and 5-minute decay times. The power spectra is found to be asymmetric with enhanced power in the high-frequency wing suggesting some contribution from waves that propagate both parallel and transverse to the field lines.
\\
\\
\noindent Acknowledgements: We are grateful for the use of SDO/AIA, STEREO and GOES data. We also thank MSRC, University of Sheffield (UK) for financial support to R.A.M for his visit to Sheffield. BWH acknowledges NASA grants NNX14AG05G and NNX14AC05G.


\end{document}